\documentclass[prd,twocolumn,nofootinbib]{revtex4-1}
\usepackage{lipsum} % placeholder text
\usepackage{multirow,xspace}
\usepackage{amsmath,amsfonts,amssymb}
\usepackage{graphicx}
\usepackage{ulem,fancyvrb}
\usepackage{hyperref}

\usepackage{cleveref}
\crefname{section}{Section}{Sections}
\crefname{equation}{Eq.}{Eqs.}
\crefname{figure}{Fig.}{Figs.}
\crefname{table}{Table}{Tables}

\usepackage[usenames,dvipsnames,svgnames,table]{xcolor}
\usepackage{xfrac}

\let\oldsqrt\sqrt
\def\sqrt{\mathpalette\DHLhksqrt}
\def\DHLhksqrt#1#2{%
\setbox0=\hbox{$#1\oldsqrt{#2\,}$}\dimen0=\ht0
\advance\dimen0-0.2\ht0
\setbox2=\hbox{\vrule height\ht0 depth -\dimen0}%
{\box0\lower0.4pt\box2}}
\newcommand{\TeV}{\ensuremath{\,\mathrm{Te\kern -0.08em V}}\xspace}
\newcommand{\GeV}{\ensuremath{\,\mathrm{Ge\kern -0.08em V}}\xspace}
\newcommand{\MeV}{\ensuremath{\,\mathrm{Me\kern -0.08em V}}\xspace}

\newcommand{\tev}{\ensuremath{\mathrm{Te\kern -0.08em V}}\xspace}
\newcommand{\gev}{\ensuremath{\mathrm{Ge\kern -0.08em V}}\xspace}

\newcommand{\NEUAffil}{Department of Physics, Northeastern University, Boston, MA 02115, USA}

\def\cl{{\cal L}}

\allowdisplaybreaks[1]

\begin{document}

\title{Reminiscences of my work with Richard Lewis Arnowitt
}
%boson discovery}

\author{Pran~Nath} 
\email[Email: ]{nath@neu.edu}
\affiliation{\NEUAffil}

\begin{abstract}

 This article contains reminiscences  of the collaborative work that Richard Arnowitt
 and I did together which stretched over many years and encompasses 
 several areas of particle theory.
 The article is an extended version of my talk at the Memorial Symposium in honor of Richard Arnowitt 
at Texas A\&M, College Station, Texas, 
September 19-20, 2014.
    \end{abstract}

\keywords{Higgs, Muon Anomalous Magnetic Moment, B-physics, Dark Matter, LHC, Supersymmetry}

\maketitle
My collaboration with Richard   Arnowitt  (1928-2014)\footnote{Obituary: Richard Lewis Arnowitt, 
 Physics Today 67(12), 68 (2014). 
Submitted by  Stanley Deser (Brandeis and Caltech),  Charles Misner (University of Maryland),
Pran Nath (Northeastern University), and  Marlan Scully (Texas A\&M and Princeton University).} 
started soon after I arrived at Northeastern University 
in 1966 and continued for many years.
In this brief article on the reminiscences of my work with Dick I review some of the more salient parts of our collaborative work 
 which includes  effective Lagrangians, current algebra, 
 scale invariance and its breakdown, the $U(1)$ problem, formulation of the first 
local supersymmetry and the development of supergravity grand unification in
 collaboration of Ali Chamseddine, and  its applications to the search for supersymmetry. 

\section{ Effective Lagrangians and  Current Algebra }

In 1964 Gell-Mann~\cite{GellMann:1964tf}
proposed that ``quark-type'' equal-time commutation relations for the vector and the axial
vector currents of weak interaction theory serve as a basis for calculations involving strongly interacting particles. 
Combined with the conserved vector current (CVC), partially conserved vector current (PCAC)
 and the soft pion approximation many successful results were obtained (see, for example, 
\cite{Weinberg:1966kf}).
However, around 1967  an important issue arose which concerned the 
breakdown of the soft pion approximation in the analysis of $\rho\to \pi \pi$
and $A_1\to \rho \pi$ where the soft pion approximation gave very poor results~\cite{Geffen,Renner}.
This problem was overcome in work with Dick and Marvin Friedman by giving up the soft pion approximation and using
the effective Lagrangian 
method to compute the mesonic vertices~\cite{Arnowitt:1967zz,Arnowitt:1969vw,Arnowitt:1969wv}.
 A number of other techniques were being pursued at that time such as  Ward identities  by
  Schnitzer and Weinberg \cite{Schnitzer:1967zzb}, phenomenological Lagrangians by
  Schwinger\cite{Schwinger:1967tc},
Wess and Zumino ~\cite{Wess:1967jq}, 
 and by Ben Lee and Nieh~\cite{Lee:1967ug} and other techniques~\cite{brown-west,Das:1967ek,Vaughn:1971tg}.
 
Here I describe briefly the approach that Dick Arnowitt, Marvin Friedman and I followed which first of all involved developing 
 an effective Lagrangian for the $\pi \rho A_1$ system but then using current algebra conditions to 
 constrain the parameters of the effective Lagrangian.
 The effective Lagrangian was a deduction from the following set of conditions:
(i) single particle saturation in computation of T-products of currents,
(ii)  Lorentz invariance, (iii)  ``spectator'' approximation, and (iv) locality which implies a smoothness assumption on the vertices. The above assumptions lead us to the conclusion
that the simplest way to achieve these constraints is via an effective Lagrangian which for T-products of 
three currents requires  writing cubic interactions involving $\pi, \rho$ and $A_1$ fields and allowing for 
no derivatives in the first -order formalism and up to one-derivative in the second order formalism.
The effective Lagrangian is to be used to first order in the coupling constants for three point functions
~\cite{Arnowitt:1967zz,Arnowitt:1969vw,Arnowitt:1969wv} and to second  order in the coupling constants for 4-point functions
and to $N-2$ order in the coupling constant for $N$ point functions. 
The second step consisted of the imposition of the constraints of current algebra, 
CVC and PCAC to determine the parameters appearing in the effective Lagrangian. 
Thus in addition to ~\cite{Arnowitt:1967zz,Arnowitt:1969vw,Arnowitt:1969wv} several other 
 applications of the effective Lagrangian method were made~\cite{Arnowitt:1969ff,Arnowitt:1971nq,Arnowitt:1969gn,Arnowitt:1969jc,Arnowitt:1971sk,Arnowitt:1971ri,Arnowitt:1969zz}.
Below we give some further details of the effective Lagrangian construction.

We begin by considering a T-product of three currents, i.e.
\begin{equation}
F^{\alpha \mu\beta} \equiv<0|T(A^\alpha_a(x) V^\mu_c(z) A^\beta_b(y))|0> \,,
\label{eff3}
\end{equation} 
For the time ordering $x^0>z^0>y^0$  \cref{eff3}  can be expanded so that 
\begin{equation}
 F^{\alpha\mu\beta}= \sum_{n,m} <0|A^\alpha_a|n><n|V_c^\mu|m><m|A_b^\beta|0> \,.
 \label{eff4}
 \end{equation}
Here the states $n$ and $m$ can only be either $\pi$ or $A_1$ mesons. 
It is clear that the matrix elements that are involved are $<0|A^\alpha_a|\pi q_1a_1>$,
$<\pi q_1a_1|V^\mu_c| \pi q_2 a_2>$ and $<\pi q_2 a_2 |A^\beta_b|0>$ and additional terms 
where $\pi$ is replaced by $A_1$.
For another time ordering , i.e., $y^0> x^0 > z^0$ one has 

\begin{equation}
 F^{\alpha\mu\beta}= \sum_{n,m} <0|A^\beta_b|n><n|A^\alpha_a|m><m|V^\mu_c|0> \,,
 \label{eff5}
 \end{equation}
where the state $n$ must be a $\pi$ or $A_1$ state. However, the state $m$ must be a $\rho$ state
to have non-vanishing matrix elements.  In addition to the above we  must also include two particle
intermediate states so that for the time ordering $x^0>y^0>z^0$ we have 

\begin{equation}
F^{\alpha\mu\beta}= \sum_{n,m,k} <0|A^\alpha_a|n,m><n,m|V^\mu_c|k><k|A^\beta_b|0> \,.
\label{eff6}
\end{equation}
Here, for example,  $n$ must be either $\pi$ or $A_1$, and $m$   must be a $\rho$ and $k$   a $\pi$ or $A_1$. 
This means that one of the particles
 in the matrix element $<n,m|V^\nu|k>$ must be  a ``spectator''. Thus one has 

\begin{align}
<\pi q_1 a_1 \rho, p_1a_3|V^\mu_c |\pi q_2a_2>&= \delta_{a_1a_2} \delta^3(\vec q_1-\vec q_2)\nonumber\\
 &\times <\rho p_1 a_3|V^\mu_c|0>\,.
\label{eff7}
\end{align}
Such contributions are required by Lorentz invariance and crossing symmetry. Indeed \cref{eff7} is the 
cross diagram
contribution to \cref{eff5}. Now the vacuum to one particle matrix elements of currents define the interpolating 
constants $F_\pi, g_A, g_\rho$ so that   one has 
 $<0|A^\alpha_a(0)|\pi q_1a_1>= F_\pi q^\alpha$,  $<0|A^\alpha(0)|A_1\sigma>= g_A \epsilon^{\alpha\sigma}$ and
 $<0|V^\mu(0)|\rho\sigma>=  g_\rho \epsilon^{\mu\sigma}$ (where we have  
 suppressed the normalizations and iso-spin factors).
 These results can be simulated by writing 
$V^\mu_a = g_\rho \tilde v^\mu_a$ and $A^\mu_a= g_A \tilde a^\mu_a + F_\pi \partial^\mu \tilde \phi_a$ 
where  the tilde fields are the in-fields. For the matrix elements of a current between two particle states, e.g.,
$<\pi q_1 a|V^\mu_c(0)|\pi q_2 b>$ we write (suppressing normalization factors)  
 $<\pi q_1 a|V^\mu_c(0)|\pi q_2 b>= \epsilon_{abc} \Delta^\mu_\lambda \Gamma^\lambda(q_1,q_2)$
 where $\Delta^\mu_\lambda$ is a $\rho$ propagator and the vertex $\Gamma^{\lambda}$ can be expanded in 
 a power series
 
 \begin{equation}
 \Gamma^\lambda(q_1, q_2)= (q_1^\lambda+q_2^\lambda) (\alpha_1 + \alpha_2  k^2+ \cdots)\,, 
\label{eff8} 
 \end{equation}
where $k^\lambda=q^\lambda_1-q^\lambda_2$. Now the two particle matrix element of $V^\mu$, i.e., 
$<\pi q_1 a|V^\mu_c(0)|\pi q_2 b>$, can be obtained if we replace the $v^\mu_c$ field in terms
of the bilinear product of two pion in-fields. Including the vacuum to one particle matrix element contribution this
 leads us to a following form for $V^\mu_c$.
 
\begin{align}
V^\mu_c(x)&= g_\rho \tilde v^\mu_c +\epsilon_{abc}
\int d^4y \Delta^\mu_\lambda(x-y)\nonumber\\
&\times  \left[\alpha_1- \alpha_2 \Box^2 + \cdots\right]  \tilde \phi_a(y)\partial^\lambda \tilde \phi_b(y)
+ \cdots\,,
\label{eff9}
\end{align}
where
the dots at the end stand for other bilinear terms that are left out. The form above guarantees 
crossing symmetry. A very similar analysis holds for the axial current. 
 For further analysis it is useful to replace the in-fields by the Heisenberg field operators  which obey the Heisenberg
 field equations. Thus we consider the $\rho$ field $v^\mu_c$ to obey the Heisenberg field equation
 
 \begin{align}
 K^\mu_\lambda (x) v^\lambda_c(x) &= g_\rho^{-1}  \epsilon_{abc} \left[\alpha_1- \alpha_2 \Box^2 + \cdots\right] \nonumber\\
&~~~~\times \phi_a(x) \partial^\mu \phi_b(x) + \cdots \,,
  \label{eff10}
  \end{align}
where $K^\mu_\lambda$ is the Proca operator $K^\mu_\lambda= (-\Box^2 + m_\rho^2) \delta^\mu_\lambda + \partial^\mu \partial_\lambda]$. Thus \cref{eff9} is now equivalent to the relation 
\begin{equation}
V^\mu_a(x) = g_\rho v^\mu_a(x).
\label{eff11}
\end{equation}
In a similar way one has 
\begin{equation}
A^\mu_a(x) = g_A a^\mu_a(x) + F_\pi \partial^\mu \phi_a(x)\,,
\label{eff12}
\end{equation}
It should be clear that \cref{eff11,eff12} are a consequence of single particle saturation assumption and not meant to 
be fundamental postulates. Thus our approach differs from the one by Lee, Weinberg and Zumino~\cite{Lee:1967iu}.
It should now be noted that \cref{eff11,eff12} are to be used  in the following way: we solve the Heisenberg equations 
and then use them to first order in the coupling constant in the computation of T-product involving three currents. 
This is equivalent to the in-field expansion of \cref{eff9}. However, due to the presence of the propagators
$\Delta^\mu_\lambda$ etc the locality of $V^\mu_c(x)$ etc is not guaranteed. Thus if we demand
that $v^\mu_a(x)$, $\pi_a(x)$ and $a^\mu_a(x)$ be local field operators, whose commutators for space-like separations 
vanish, then this condition can be guaranteed if we require that the sources for the fields $v^\mu_c$ etc arise from 
a Lagrangian with interactions that are cubic in the fields. Thus the arguments laid out above lead us to an effective
Lagrangian of the form 

\begin{equation}
\cl_{eff}^{(3)}= \cl_0  + g \cl_3
\label{eff13}
\end{equation}
where $\cl_0=\cl_{0\pi} + \cl_{0\rho} + \cl_{0A_1}$. 
We note that we arrived at \cref{eff13} purely from the conditions of (i) single particle saturation of T -products,
(ii) spectator approximation, (iii) Lorentz invariance and crossing symmetry and , (iv) locality. The current
algebra constraints, i.e., current algebra commutation relations, CVC and PCAC  have played no role in the analysis thus far.\\

The analysis above can be  extended to higher point functions. For instance, for the computation
of the four -point function,
\begin{equation}
 F^{\alpha\beta\mu\nu}(x,y,z,w)\equiv <T(A^\alpha(x) A^\beta (y) V^\mu(z) V^\nu(\omega))>\,,
\end{equation}
 the implementation of the single particle saturation requires that we include 
the following set of contributions, and  (i) diagrams where we have a cascade of three point vertices, and (ii) diagrams where we
have  four point vertices. The diagrams of type (i) arise from ${\cal L}_3$, and  diagrams of type (ii) arise from ${\cal L}_4$.  The Lagrangian
 \begin{equation}
 {\cal L}_{eff}^{(4)}= \cl_0+ g {\cal L}_3 +g^ 2{\cal L}_4\ ,
  \end{equation}
is to be used to the first non-vanishing order, i.e., order $g^2$,  discarding disconnected diagrams to compute the  T product of four  currents. 
It is now straightforward to extend the analysis to the computation of T products of N-point functions.
Thus using the same principles as above,  the analysis of N-point functions involves using an interaction 
Lagrangian ~\cite{Arnowitt:1969ff,Arnowitt:1971nq}

\begin{equation}
{\cal L}_{eff}^{(N)}=\cl_0+ 
g {\cal L}_3+ g^2{\cal L}_4+ \cdots+ g^{k-2} {\cal L}_k+ \cdots + g^{N-2} {\cal L}_N\ . 
\end{equation}
The coupling in  ${\cal L}_k$ will be viewed as ${\cal O} (g^{k-2})$.  Further, the computation of 
an N -point function is then done using the effective  Lagrangian  to order $N-2$ in the 
couplings. 
The effective Lagrangian techniques described above  
have a much larger domain of validity than the specific example of the mesonic system being discussed here.

{\it Current Algebra Constraints:} 
After ensuring that the constraints of single particles saturation, spectator approximation, Lorentz invariance and locality can be embodied by 
writing an effective Lagrangian we impose constraints of current algebra.
As mentioned earlier  these consist of 
(i) equal-time commutation relations on the densities\footnote{The imposition of the absence of  q-number Schwinger term 
gives the first Weinberg Sum rule:
$g_\rho^2/m_\rho^2 = g_A^2/m_A^2 + F_\pi^2$.}, (ii) CVC, and (iii) PCAC. 
  The $\pi-\rho-A_1$ effective Lagrangian allowed one
 to compute $\pi \rho A_1$ processes  without the soft pion approximation and 
 get results consistent with data. As discussed above the
 technique of effective Lagrangian allows one to  obtain Lagrangians obeying current algebra constraints
  for higher points functions. Thus for  $SU(2)\times SU(2)$    
  current algebra constraints the effective Lagrangian method was used not only for the processes $\rho\to \pi \pi$ , $A_1\to \pi \rho$ but also to give the 
 first analysis of $\pi\pi\to \pi \pi$ scattering using hard pion current algebra ~\cite{Arnowitt:1969gn}. 
 The effective Lagrangian method was then extended to include $SU(3)\times SU(3)$ current algebra constraints 
which allowed an analysis of the $K\ell_3$  and  $\pi K$ scattering ~ \cite{Arnowitt:1969jc,Arnowitt:1971sk,Arnowitt:1971ri,Arnowitt:1970wt}. In  a later work ~\cite{Arnowitt:1970wt} the current algebra constraints were also applied to the Veneziano model.\\

{\it Effective  vs Phenomenological Lagrangians:}
The effective Lagrangian technique used in 
\cite{Arnowitt:1967zz,Arnowitt:1969vw,Arnowitt:1969wv, Arnowitt:1969ff,Arnowitt:1971nq,Arnowitt:1969gn,Arnowitt:1969jc,Arnowitt:1971sk,Arnowitt:1969zz,Arnowitt:1971ri}.
 is different from the works of Schwinger~\cite{Schwinger:1967tc},
Wess and Zumino~\cite{Wess:1967jq}, 
 and of Ben Lee and Nieh~\cite{Lee:1967ug} 
   which were phenomenological
 Lagrangians.
 In phenomenological Lagrangian, one starts by constructing Lagrangians which have $SU(2)\times SU(2)$ 
 or $SU(3)\times SU(3)$  invariance.
 The invariance is then broken by additional terms which are introduced by hand. 
 In the effective Lagrangian approach no a priori assumption was made regarding the type of symmetry breaking, chiral or ordinary. The current algebra  constraints alone determine the nature of symmetry breaking. 
 What we found was that for hard meson current algebra with single meson dominance of the currents and of the
 $\sigma$ commutator, the chiral symmetry breaking was broken only by~\cite{Arnowitt:1971nq} 
  $(3,3^*)+ (3^*,3).$
 This type of 
 breaking had been proposed by Gell-Mann, Oakes and Renner~\cite{GellMann:1968rz}.

{\it  The Axial Current Anomaly:}
 Beginning in 
  1967 one of the big puzzles related to the Veltman theorem~\cite{veltman}
   which was that in the soft
 pion approximation the pion decay into two $\gamma's$ vanished.
 It was generally held that a possible source of this problem could be that the soft pion approximation was breaking down due to a very rapid variation of the matrix elements as we went off  the pion mass-shell.
 However, in a paper in 
 1968 we discovered~\cite{afn-pi0decay}
  that hard pion analysis also gave a vanishing $\pi^0 \to 2 \gamma$ decay.
 This lead us to propose a modification of the PCAC condition by introducing an axial current anomaly
  which  exists even in the chiral limit~~\cite{afn-pi0decay}, i.e., we proposed   
 $\partial_\mu A^\mu_a = F_a m_a^2 \phi_a + \lambda d_{abc} \epsilon_{\mu\nu\alpha\beta} F_b^{\mu\nu} F_c^{\alpha\beta} +
 \lambda' \epsilon_{\mu\nu\alpha\beta} F^{\mu\nu}_a \phi^{\alpha\beta}\,,$
where $a,b,c=1\cdots 8$.
With the above modification a number of other decays were also computed such as $\eta\to 2 \gamma$ and
$\rho^0 \pi^0 \gamma$.  The axial anomaly was simultaneously computed from triangle loops with fermions by 
 Bell and Jackiw ~\cite{Bell:1969ts} and  by Adler~\cite{Adler:1969gk}.

{\it Scale Invariance and scale breaking:}
The scaling~\cite{Bjorken:1968dy} observed in electro-production data lead  to the hypothesis that physical laws are scale-invariant at high energies~\cite{Mack:1968zz,Kastrup:1966zzc,Wilson:1969zs}.  
While such a hypothesis may hold at high energy, it is certainly violated at low and intermediate
energies as physics is not scale-invariant there. Thus at low and intermediate scales dimensioned parameters  
such as masses appear. Supposing that scale invariance is of fundamental significance then 
it is of relevance to ask the manner of its breakdown. Several works had  tried to approach this problem
in the early seventies ~\cite{zumino-brandeis,Carruthers:1971vz,DeAlwis:1970ig,Isham:1971gm}.
In our analysis of scale invariance and its breakdown  the stress tensor and its trace  play a central role. 
Out approach to scale invariance and its breakdown was parallel to our approach to current algebra~\cite{Nath:1973yf,Friedman:1973px,Arnowitt:1973iq}.
 Thus in the analysis of the current algebra constraints we assumed that the vector current was dominated by a vector
 spin 1 particle and the axial current by  an axial vector and a pseudo-scalar meson. In an  analogous fashion
 we assumed that the stress tensor was dominated by $J^P=2^+, 0^+$ mesons in a new field current identity. 
 Applications were made in the deduction of light-cone algebra, deep-inelastic scattering~\cite{Arnowitt:1973iq} and 
 $e^+e^-$ annihilation at intermediate energies ~\cite{Friedman:1973px}.

\section{ The $U(1)$ Problem}
The $U(1)$ problem relates to the fact that the ordinary $U(3)\times U(3)$ current algebra leads to 
the ninth pseudo-scalar meson being  light ~\cite{glashow,Weinberg:1975ui}: 
$m_{\eta'} < \sqrt 3 m_\pi \,.$
Formally  a resolution was proposed by t'Hooft ~\cite{'tHooft:1973jz}
who showed that the instanton solution to the Yang-Mills 
theory provides a contribution to the $\eta'$ mass.
However, the t'Hooft solution is inadequate as it does not make contact with the quark-antiquark
annihilation of the singlet pseudo scalar into gluons
\cite{DeRujula:1975ge}.
Further, Witten\cite{Witten:1979vv}
showed that a resolution of the $U(1)$ anomaly 
arises in the $1/N$ expansion of QCD. 
Thus the $\eta'$ is massless in the $N\to \infty$ limit but significant
non-zero contributions arise from terms which are $1/N$ smaller than the leading terms and split $\eta'$ from the
octet.
We examined the problem from an effective Lagrangian view point.
We introduced a Kogut-Susskind ghost field $K^\mu$ and constructed a closed form solution for the 
 effective Lagrangian\cite{Nath:1979ik,Arnowitt:1980ne,Nath:1980nf,Arnowitt:1981kh}.
\begin{equation}
{\cal L}= {\cal L}_{CA} + \frac{1}{2C} (\partial_\mu K_\nu)^2 + G \partial_\mu K^\mu - \theta \partial_\mu K^\mu\ , 
\end{equation}
which gives a complete description of the interaction of the field $K^\mu$ with the mesonic fields. 
Thus
 $G$ is a function which depends on the spin zero and spin one mesonic fields, $\theta$ is the strong CP violating  parameter of $QCD$, and $C$ is the strength of the topological 
charge $i<T(K_\mu K_\nu)>$ = $- C\eta_{\mu\nu} /q^2+ \cdots$.
  Ignoring spin one fields, $G$ is  determined to be of the form given by Rosenzweig et.al.~\cite{Rosenzweig:1979ay}, 
  %\begin{equation}
$G= \frac{1}{2} \left[ {\rm ln det} \xi - {\rm ln det} \xi^\dagger\right]$,
%\end{equation}
where $\xi= (u_a + i v_a) \lambda_a$ and where $u_a$ and $v_a$ are scalar and pseudo-scalar densities. 
Using the effective Lagrangian which includes the effect of the $U(1)$ anomaly,  we found a  sum rule
of the form~\cite{Arnowitt:1980ne,Nath:1980nf}
\footnote{$F_{ab}$ are defined through divergence of the axial current so that
$$\partial_\mu A^\mu_a \supset F_{ab} \mu_{bc} \chi_c + \sqrt{\frac{2}{3}} N_\ell \delta_{a9} \partial_\mu K^\mu.$$}

\begin{align}
&(F_{88} + \sqrt 2 F_{98})^2 m_\eta^2 +
(F_{89} + \sqrt 2 F_{99})^2 m_{\eta'}^2 \nonumber\\
&~~~~~~~~~~~~~= 3 m_\pi^2 F_\pi^2 + \frac{4}{3} N_f^2 \left(\frac{d^2 E}{d\theta^2}\right)^{N_f=0}_{\theta=0}\,,
\end{align}
where  $N_f$  is the number of light quark flavors.
If one ignores the first and the last terms, sets $F_{89}=0$, and let $F_{99}\to \sqrt{N_f/6} F_\pi$ $(N_f=3)$, one finds the Weinberg result~\cite{Weinberg:1975ui}
\begin{equation}
m_{\eta'} < \sqrt 3 m_\pi\,. 
\end{equation}
Further, in the limit $m_\pi=0=m_\eta$, 
$F_{89}=0$ and $F_{99}\to \sqrt{N_f/6} F_\pi$ one finds Witten's result \cite{Witten:1979vv}
\begin{equation}
m_{\eta'}^2\to \frac{4N_f}{F_\pi^2} \left(\frac{d^2 E(\theta)}{d\theta^2}\right)^{N_\ell=0}_{\theta=0}.
\end{equation}
In addition to the work of \cite{Nath:1979ik,Arnowitt:1980ne,Nath:1980nf,Arnowitt:1981kh,Rosenzweig:1979ay} a Lagrangian formulation including the 
$U(1)$ axial anomaly was given by
DiVecchia and Veneziano~\cite{DiVecchia:1980ve}.
Witten~\cite{Witten:1980sp} has shown that these  Lagrangian formulations  which include the effect of the
$U(1)$ anomaly and solve the $\eta'$ puzzle are consistent with  the large N chiral
dynamics.

     \section{Local Supersymmetry}
It was in 1974 when I  was at the XVII ICHEP Conference in London that I first became interested in supersymmetry.
On my  return to Boston
I talked to Dick to work in this area. At that time SUSY was a global symmetry, and we thought that if it is a fundamental symmetry
it ought to be a  local symmetry.  Very quickly we realized that {\it gauging of supersymmetry requires bringing in gravity}, and we  thought that
the direct course
of action was to extend the geometry of Einstein gravity to  superspace geometry.
This lead to  the formulation of gauge supersymmetry~\cite{Nath:1975nj}
based on a single  tensor superfield $g_{\Lambda\Pi}(z)$
in superspace consisting of bose and fermi co-ordinates, i.e., 
$z^\Lambda=(x^\mu, \theta^{\alpha i})$  where $x^\mu$ are the bose co-ordinates of ordinary space-time
and $\theta^{\alpha i}$ are anti-commuting fermi co-ordinates.
 We considered a line element of the form $ds^2=dz^\Lambda g_{\Lambda\Pi}(z) dz^\Pi$ and required its
invariance under the general co-ordinate transformations in superspace $z^{\Lambda}= z^{'\Lambda} + \xi^\Lambda(z)$ which 
leads to the transformations of the superspace metric tensor of the form

\begin{equation}
\delta g_{\Lambda\Pi}(z)= g_{\Lambda\Sigma} \xi^\Sigma_{,\Pi} + (-1)^{\Lambda+ \Lambda\Sigma} \xi^\Sigma_{,\Lambda}
\xi_{\Sigma\Pi} + g_{\Lambda\Pi,\Sigma} \xi^\Sigma\,,
\label{deltagAB}
\end{equation}
where $(-1)^{\Lambda}= 1(-1)$ when $\Lambda$ is bosonic (fermionic) etc.
One may also introduce a supervierbein so that
\begin{equation}
g_{\Lambda \Pi}(z) = V^A_{~~\Lambda}(z) \eta_{AB} (-1)^{(1+B) \Pi} V^B_{~~\Pi} (z)
\end{equation}
where $\eta_{AB}$ is a tangent space metric so that 

\begin{equation}
\eta_{AB} = \left(\begin{matrix} \eta_{mn} & 0 \\ 
0 & k \eta_{ab} \end{matrix} \right)
\label{gb1}
\end{equation}
where $\eta_{mn}$ is the metric in bose space and $\eta_{ab}$ is 
the metric in fermi space, so that $\eta_{ab}= -(C^{-1})_{ab}$ where $C$ is the charge conjugation matrix.
In \cref{gb1}~$k$ is an arbitrary parameter.  
Global supersymmetry transformations are generated by $\xi^\Lambda$ of the form

\begin{equation}
\xi^\mu =i \bar\lambda  \gamma^\mu \theta, ~~~ \xi^\alpha= \lambda^\alpha\,,
\end{equation} 
where $\lambda^\alpha$ are constant infinitesimal anti-commuting parameters. For the global supersymmetry
case the metric that keeps the line element invariant is given by 
\begin{align}
g_{\mu\nu} &= \eta_{\mu\nu}\,, \nonumber\\
g_{\mu\alpha} &= -i(\bar\theta \gamma^m)_{\alpha} \eta_{m \mu}\,,\nonumber\\
g_{\alpha \beta}&= k \eta_{\alpha\beta} + (\bar \theta \gamma_m)_{\alpha} (\bar \theta\gamma^m)_\beta\,.
\label{global}
\end{align}
 To set up an action principle in superspace required defining a superdeterminant. 
 In work with Bruno Zumino~\cite{Arnowitt:1975xg} it was shown that
for a matrix $M_{AB}$ with  bosonic and fermionic components $\{M_{\mu\nu}, M_{\mu\alpha}, M_{\alpha\mu}, M_{\alpha\beta}\}$ where $M_{\mu\nu}$ and $M_{\alpha\beta}$ are bosonic and $M_{\mu \alpha}$ and
$M_{\alpha\mu}$ are fermionic quantities,  $det(M)$ is given by

\begin{equation}
det M= (det M_{\mu\nu}) det(M^{-1})^{\alpha\beta}\,.
\end{equation}
This also then gives $\sqrt {(-g)}= [-(det g_{\mu\nu}) (det g^{\alpha\beta})]^{1/2}$. 
 Using the above one can then set up an action principle in superspace so that 
 
 \begin{equation}
 A =\int d^8z \sqrt{-g} R \,,
 \label{A1}
 \end{equation}
 where  $R$ is a superspace curvature scalar defined by $R= (-1)^a g^{AB} R_{BA}$.
 A more general action than the one in \cref{A1} can be gotten by including an additional term, i.e.,
 $2\lambda \int d^8z \sqrt{-g}$,   which 
 leads to the field equations in superspace to read $R_{AB} = \lambda g_{AB}.$
Some very encouraging features emerged. From the superspace transformations we were able to recover both the
Einstein gauge invariance and the Yang-Mills gauge invariance which appeared to be quite remarkable.
 The metric superfield  $g_{AB}$ contains many fields.  Thus 
$g_{\mu\nu}(z) ={ g_{\mu\nu}(x)} + \cdots$, $g_{\mu\alpha}(z)=   {\bar \psi_{\mu\alpha}(x)} + (\bar \theta M_x)_\alpha {B_\mu^x} + \cdots$, 
$g_{\alpha\beta}(z)= (\eta F(x))_{\alpha\beta} +(\bar\theta M_x)_{[\alpha} \bar \chi^x_{\beta]}(x) + \cdots$.
 Here $g_{\mu\nu}$ is the metric in ordinary 
bose space and contains the spin 2 graviton field, $\psi_{\mu \alpha}$ is a spin $3/2$ field, $B_\mu$ is  a vector field and $F(x)$ and $\chi_\beta$ were spin 0 and spin 1/2 fields. The implications of this theory was examined in several
works  \cite{Arnowitt:1975bd, Arnowitt:1976nm,Nath:1976rk,Arnowitt:1978me,Nath:1978av,Arnowitt:1978gz,Arnowitt:1979cz,Arnowitt:1979jj}.
  
 {\it Supergravity and Gauge Supersymmetry}:
   While gauge supersymmetry was the first realization of  a local supersymmetry, 
 its field content is rather complicated. 
     Further developments in this field occurred 
     via formulation of local supersymmetry involving just the 
spin 2 and spin 3/2 fields~\cite{Freedman:1976xh,Deser:1976eh} (for a review see~\cite{VanNieuwenhuizen:1981ae}),
i.e., supergravity.
The question then is what is the connection of gauge supersymmetry
to  supergravity. 
In ~\cite{Nath:1976ci,Arnowitt:1978jq,Nath:1979yn} we showed 
 that   supergravity  
can be recovered from gauge supersymmetry if one discards all other fields in the metric $g_{AB}(z)$ 
except the spin 2 and spin 3/2
fields and considers the $k\to 0$ (where $k$ is defined in \cref{global})  limit of gauge supersymmetry.

Thus to recover supergravity from gauge supersymmetry
 we construct the metric only in terms of spin 2 and spin 3/2 fields. 
 Specifically we want to construct $g_{AB}$ depending on  the  fields $e^m_{~\mu}(x)$ and $\psi^\alpha_\mu(x)$ 
 and find $\xi^A$ so that the transformation equation \cref{deltagAB} leads correctly to the supergravity
 transformations for $e^m_{~\mu}(x)$ and $\psi_\mu(x)$ so that 
 \begin{align}
 \delta e^m_{~\mu}&= i \bar \psi_\mu(x) \gamma^m \lambda(x)\,,\nonumber\\
 \frac{1}{2} \delta \psi_\mu(x) &= (\partial_\mu + \Gamma_\mu) \lambda(x),,
 \label{sg1}
 \end{align}
 where $\lambda(x)$ are the transformation parameters and $\Gamma_\mu$ is to be determined.
 This is to be done by using  \cref{deltagAB} order by order in $\theta$ by the process of 
 gauge completion developed in  ~\cite{Nath:1976ci,Arnowitt:1978jq,Nath:1979yn}.
 Using this procedure one finds ~\cite{Nath:1976ci,Arnowitt:1978jq,Nath:1979yn} 
 \begin{align}
 \xi^\mu(z) &= i\bar \lambda(x)\gamma^\mu\theta + \frac{1}{2} (\bar \psi_m \gamma^\mu \theta)
 (\bar \lambda \gamma^m\theta) + \Delta \xi^\mu(z)\,,\nonumber\\
 \xi^\alpha(z)& = \lambda^\alpha(x) - \frac{i}{2}  \psi_m^{~\alpha} \bar \lambda \gamma^m \theta 
 -\frac{1}{4} \psi_r^{~\alpha} (\bar \psi_m \gamma^r \theta) (\bar \lambda \gamma^m \theta)\nonumber\\
 &- i (\Gamma_m \theta)^\alpha (\bar \lambda \gamma^m \theta) + \Delta \xi^\alpha(z)\,,\nonumber\\
    g_{\mu\nu}(z) &= g_{\mu\nu}(x) + i \bar \psi_{(\mu} \gamma_{\nu)} \theta -  i\bar \theta \gamma_{(\mu} \Gamma_{\nu)} \theta\nonumber\\
&   - (\bar \psi_\mu \gamma^m \theta)
   %\nonumber\\ 
  (\bar \psi_\nu \gamma_m  \theta) + \delta g_{\mu\nu} \,,\nonumber\\
   g_{\mu\alpha}(z)&= - i(\bar \theta \gamma_m)_{\alpha} e^m_{\mu} + (\bar\psi^\mu \gamma_m \theta) 
 (\bar\theta  \gamma^m)_\alpha + \Delta g_{\mu\alpha}\,,\nonumber\\
 g_{\alpha\beta}(z)&= k \eta_{\alpha\beta} +  (\bar\theta \gamma_m)_\alpha (\bar \theta \gamma^m)_\beta
  + \Delta g_{\alpha\beta}\,. 
  \label{trans-eqs}
 \end{align}
 Here $\Delta \xi^\mu, \Delta \xi^\alpha$ etc are quantities that depend on $k$ and also contain terms 
 $O(\theta^3)$ and higher.  An important result that emerges is that the gauge completion procedure
using  \cref{deltagAB} determines the vierbein affinity to be that of supergravity, i.e., 

\begin{equation}
\Gamma_\mu= \frac{i}{4} \sigma_{rs} \omega_\mu^{~rs}\,,
\end{equation} 
  where $\omega_\mu^{~rs}$   correctly includes the supergravity torsion. In the $k\to 0$ 
  \cref{trans-eqs} give 
   the correct supergravity transformation equations as well as the correct $g_{\Lambda \Pi}$  up to
   ${\cal{O}}(\theta^2)$.
   Further the dynamical equations 
  of  gauge supersymmetry $R_{\Lambda \Pi}=0$ (setting $\lambda=0$) produce correctly 
  the dynamical equations of supergravity. We note here that the integration of ~\cref{deltagAB}
  beyond linear order in $\theta$ requires  use of on-shell constraints, i.e., they can be integrated if 
  we impose field equations. Integration off the mass-shell requires  Breitenlohner  fields~\cite{Breitenlohner:1976nv}  which allows  gauge completion without use of field equations \cite{Arnowitt:1978jq}.

 Geometrically the connection between  gauge supersymmetry and supergravity is the following:
 Gauge supersymmetry is a geometry with the tangent space group  $OSp(3,1|4N)$  while the tangent
 space group of supergravity geometry is  $O(3,1)\times O(N)$.  
In the limit $k\to 0$ the geometry of gauge supersymmetry contracts to the supergravity geometry.
The contraction produces the desired torsions ~\cite{Nath:1979yn}
needed in  the superspace formulation of supergravity
~\cite{Wess:1977fn,Brink:1978iv,Brink:1978sz} and  the tangent space group  $OSp(3,1|4N)$ 
of gauge supersymmetry  reduces ~\cite{Nath:1979yn} to the tangent space group  $O(3,1)\times O(N)$  of supergravity. Thus the
the $k\to 0$ limit of the geometry
of gauge supersymmetry correctly produces the  supergravity geometry in superspace.     
     
\section{Gravity Mediated Breaking and Supergravity Grand Unification}
 This phase of research with Dick involves Ali Chamseddine.
 In 1980 I was on sabbatical leave at CERN and it was sheer good luck that I met Ali there. 
  Ali was aware of the work that Dick and I had done on local supersymmetry since at the suggestion of 
  his thesis advisor Abdus Salam, he had worked on gauge supersymmetry which appears as a part of his
Ph.D. thesis at Imperial College, London~\cite{Chamseddine:2004ty,thesis}. 
 Beyond that he had worked on supergravity as a gauge theory of supersymmetry~\cite{Chamseddine:1976bf}.
  At the time I met Ali, Dick and I were looking for a research associate on our NSF grant and we thought that Ali 
  would be a good fit for us  because of our common interests in supersymmetry 
   and so after  a conversation with Dick, 
  I made an offer to Ali to visit Boston after culmination of his Fellowship period at CERN.
  Ali arrived in Boston in January of 1981.  He had recently finished
  a work on the coupling of $N=4$ supergravity to $N=4$ matter~\cite{Chamseddine:1980cp} and was actively working on
  interacting supergravity in ten-dimensions and its compactification to a four -dimensional theory. 
  After  his work on 10-dimensional supergravity  was finished in the beginning of the Fall 1981~\cite{Chamseddine:1981ez}, our interests converged 
  on model building within $N=1$ supergravity framework.  $N=1$ supersymmetry has been shown to have 
  many desirable properties including the fact that it provided a technical solution to the gauge hierarchy 
  problem. However, 
 at that time there were no acceptable  supersymmetry based  particle physics models where one could break 
 supersymmetry spontaneously in a phenomenologically viable way.

As mentioned our analysis started in  the beginning of the Fall of 1981. At that time only the most general coupling of one chiral field
with supergravity was known through the 1979 work of Cremmer etal~\cite{Cremmer:1978hn}. However, the construction of 
a particle physics model required extension to an arbitrary number of chiral fields.  Thus the first task
was to construct a Lagrangian with an arbitrary number of chiral
fields which couple to an adjoint representation of a gauge group and to $N=1$ supergravity.
This analysis was rather elaborate and took us up to the early spring of 1982 to complete. 
The Lagrangian showed some very interesting features in that there were terms in the scalar potential
which were both positive and negative and thus an opportunity existed of their cancellation 
after spontaneous breaking of supersymmetry which was a very desirable feature for the 
generation of a viable model. Although we had the couplings in the  early spring of 1982, we 
did not publish them immediately since we were after construction of a realistic supergravity 
grand unified model. The $N=1$ supergravity couplings were published later in the Trieste Lecture
Series titled  {\it ``Applied N= 1 Supergravity''}~~\cite{Nath:1983fp}. Our analyses to be published 
later in the Summer and Fall of 1982~\cite{Chamseddine:1982jx,Nath:1982zq,Nath:1983aw}
were based on the supergravity couplings contained in ~\cite{Nath:1983fp}.
(The supergravity couplings with an arbitrary number of chiral fields were independently obtained
   by Cremmer etal  \cite{Cremmer:1982wb,Cremmer:1982en}  and also  by Bagger and Witten~\cite{Bagger:1982fn}).
 In our attempt to construct the supergravity  grand unified
model one of the phenomena we noticed concerned the lifting of the degeneracy after spontaneous breaking of the
GUT symmetry. Thus in a globally supersymmetric theory the breaking of $SU(5)$ leads to three possible 
vacua which have the vacuum symmetry given by $SU(5)$, $SU(4)\times U(1)$ and $SU(3)\times SU(2)\times U(1)$
which are degenerate. For the supergravity case we found  that this degeneracy was lifted by gravitational interactions.
This phenomena was also observed by Weinberg \cite{Weinberg:1982id}.

However, to construct a realistic grand unified model we needed to break the $N=1$ supersymmetry and 
grow mass terms for the squarks and the sleptons which were  large enough to have escaped detection
in current experiment.  For the breaking of supersymmetry we utilized the superHiggs effect
where the superHiggs field develops a vacuum expectation value which is ${\cal O}(\rm M_{\rm Pl})$.
However, a superHiggs field could not be allowed to interact with the matter fields in the superpotential
directly as that would lead to Planck size masses for the matter fields. For this reason it was necessary to
create two sectors, one where only quarks, leptons and Higgs fields reside and the other sector where
the superHiggs field resides. In this case the only coupling that exists between the observed or the  visible sector
and the superHiggs or the  hidden sector was through gravitational interactions.  In this way we could generate
soft terms in the visible sector which could be of the electroweak size. Thus we assumed the superpotential
to be of the form 
$W=W_1+W_2$ where $W_1$ contains only matter and Higgs fields and $W_2$ only the superHiggs field $z$.
Assuming  $W_2=m^2f(z)$ one finds that the soft terms of size ${\cal O}(m^2/M_{\rm Pl})$ grow 
in the visible sector after spontaneous breaking of supersymmetry in the hidden sector. Thus if $m$ is of 
size the intermediate scale, i.e.,   ${\cal O}(10^{11})$ GeV, one finds that soft terms in the visible sector are 
size the electroweak scale. The intermediate scale of ${\cal O}(10^{11})$ GeV could arise from a strongly 
interacting gauge group in the hidden sector.

 The mechanism discussed above avoids the appearance of Planck size masses in the soft sector. 
  However, since we were working with a grand unified theory where heavy GUT fields with 
  masses of size ${\cal O}(M_G)$ appear, it was possible that the soft terms could be size
  ${\cal O}(M_G)$. Our analysis of ~\cite{Chamseddine:1982jx,Nath:1982zq}  (see also ~\cite{Hall:1983iz})
  showed that the soft terms are indeed independent of $M_G$. 
   In the analysis of ~\cite{Chamseddine:1982jx}  
  the generation of soft terms was also shown to lead to the breaking of the electroweak
   symmetry resolving a long standing problem of the Standard Model where the breaking is 
   induced by the assumption of a tachyonic mass term for the Higgs boson.
 The preceding discussion shows that our efforts were successful 
  and we were able to formulate the first 
   phenomenologically viable supergravity grand unified model with gravity mediated breaking
   \cite{Chamseddine:1982jx,Nath:1982zq,Nath:1983aw} which lead to several further works involving Ali, Dick, and myself
     \cite{Nath:1982zq,Arnowitt:1983ah,Nath:1983fp,Chamseddine:1983eg,Nath:1983iz,Yuan:1984ww,Arnowitt:1985iy,Nath:1985ub,Chamseddine:1986ee}.  A history of the development of SUGRA GUT and of these collaborative works are
     discussed in several reviews \cite{Arnowitt:2012gc,Chamseddine:2000nk,Chamseddine:2004ty}.
     Subsequent to our work of  \cite{Chamseddine:1982jx} a number of related  papers appeared in a short
     period of time. A partial list of these  is given in ~\cite{Barbieri:1982eh,Ibanez:1982ee,Nilles:1982mp,Ellis:1982wr,Weinberg:1982tp,Ibanez:1983wi,Ibanez:1983di,Ibanez:1984vq,AlvarezGaume:1983gj,Soni:1983rm}.

 The 1982 works created a new direction of research where the electroweak physics testable at colliders and in 
 underground experiment could be discussed within the framework of a UV complete model. 
    Specifically testable predictions of supergravity models    include  electroweak loop corrections to precision parameters such as $g_\mu-2$, sparticle signatures at colliders,  proton decay and dark matter. 
 Several of these topics 
 were worked on in a series of papers involving Ali, Dick and myself~\cite{Nath:1982zq,Arnowitt:1983ah,Nath:1983fp,Chamseddine:1983eg,Nath:1983iz,Yuan:1984ww,Arnowitt:1985iy,Nath:1985ub,Chamseddine:1986ee}. Further collaborative 
 work on these topics between Dick and I  continued even after Ali left Northeastern University and moved on 
 to work on strings and on non-commutative geometry.       
     Below we give further details of  some of the implications of  supergravity unified models.
  
As mentioned earlier a remarkable aspect of supergravity grand unified theory is that soft breaking parameters can induce
breaking of the electroweak symmetry~\cite{Chamseddine:1982jx} and an
attractive mechanism for this  is via radiative breaking
 \cite{Ibanez:1982ee,Ellis:1982wr,Ibanez:1983wi,AlvarezGaume:1983gj,Arnowitt:1992qp} using renormalization group 
 evolution~\cite{Martin:1993zk} (for a review see
 ~\cite{Ibanez:2007pf}).   The radiative electroweak symmetry breaking must, however, be arranged to
 preserve  color and charge  conservation~\cite{Frere:1983ag,Claudson:1983et,Drees:1985ie}. 
 The radiative electroweak symmetry breaking can  be used to determine the  Higgs mixing parameter $\mu$  except for 
 its  sign.
 The simplest SUGRA model that emerges is the one with universal soft breaking at the grand unification
 scale which can be parameterized  at the electroweak scale by 

    \begin{equation}
   m_0, m_{1/2}, A_0, \tan\beta, {\rm sign}(\mu): ~~{\rm mSUGRA}
\end{equation}
where $m_0$ is the universal scalar mass, $m_{1/2}$ is the universal gaugino mass, $A_0$ is the universal trilinear 
coupling, $\tan\beta= <H_2>/<H_1>$ where  $H_2$ gives mass to the up quarks and $H_1$ gives mass to the down
quarks and leptons, and sign($\mu$) is the sign of $\mu$ which is  not  determined by radiative 
electroweak symmetry breaking. 
For more general choices of the Kahler potential and of the gauge kinetic energy function, SUGRA models
with non-universalities are obtained~\cite{Soni:1983rm,Drees:1985bx,Ellis:1985jn,Nath:1997qm}.
Some early work on the signatures of  supergravity models  can be found in 
~\cite{Weinberg:1982tp,Arnowitt:1983ah,Chamseddine:1983eg,Nath:1983iz,Dicus:1983fe,Dicus:1983cb,Frere:1983dd,Ellis:1983er,Ellis:1983wd,Polchinski:1983zd,Grinstein:1983ky,Claudson:1983cr}.
 
{\it Supersymmetric electroweak corrections to $g_\mu-2$:}
It was realized early on ~\cite{Yuan:1984ww} (see also~\cite{Kosower:1983yw})
that the supersymmetric electro-weak corrections to the  anomalous magnetic moment of the muon in supergravity unified models could be of the same size as the electroweak corrections arising from the Standard Model.
Specifically it was shown that the supersymmetric electroweak corrections
can be substantial for low lying charginos, neutralinos and smuon states circulating in the loops.
This result was helpful 
when the Brookhaven
experiment E821 was being conceived.
The current data from the  Brookhaven experiment E821~\cite{Bennett:2006fi}  which measures
\(a_\mu = \frac{1}{2}(g_\mu-2)\) shows a deviation from the Standard Model prediction
     ~\cite{Hagiwara:2011af,Davier:2010nc}
at the $3\,\sigma$ level, i.e., it gives 
$\delta a_\mu = (287 \pm 80.)\times10^{-11}$. 
It remains to be seen if the observed deviation will survive in future improvement of experiment and
also further improvement in the analysis of the hadronic correction.

{\it Supersymmetric signals:}
Subsequent to the development of the supergravity grand unification, its possible signatures at colliders 
were investigated. One of the main focus was on jets, leptons and  missing energy signals~\cite{Chamseddine:1983eg,Nath:1983iz,Nath:1983fp}.
Initially the analyses were for the on-shell decays of the W and Z boson ~\cite{Chamseddine:1983eg,Nath:1983iz,Dicus:1983cb,Baer:1986dv,Nath:1983fp} where an on-shell  $W$ decays  via the  chain $W^{\pm} \to \chi_1^{\pm} \chi_2^0$ with the further decays $\chi_1^-\to e^- \bar \nu \chi_1^0$ and $\chi_2^0\to \ell^+\ell^- \chi_1^0$.  
This leads to a trileptonic signal $\ell_1\ell_2 \bar \ell_2$ plus missing transverse energy $E_T$. These on-shell 
$W$ decay analyses were limited  by the  constraint  $M_{\chi_1^{\pm}} + m_{\chi_2^0} < M_W$.
However, in~\cite{Nath:1987sw,Arnowitt:1987pm}
the analysis was
extended to decays when $W$ and $Z$ are off-shell. Here it was shown that strong leptonic signals can arise
even when $W$ and $Z$ are off-shell and such signals 
 are now some  of the primary modes of discovery for the supersymmetric particles.
 Leptonic signals were further discussed in several later works~\cite{Baer:1995va,Barger:1998wn,Accomando:1998cv}.  
 A variety of other signatures of SUGRA models were discussed in several early reports
 on  supersymmetric signatures ~\cite{Nath:1983fp,Nilles:1983ge,Haber:1984rc} 
 and more intensely  in the SUGRA Working Group 
Collaboration Report \cite{Abel:2000vs}.

{\it Sparticle spectrum:}
After the LEP data came out  which showed that the extrapolation of gauge coupling constants 
was not  consistent with a non-supersymmetric grand unification $SU(5)$ but was consistent with a supersymmetric
one, we found it appropriate to compute the sparticle spectrum in a supergravity unified model using 
renormalization group evolution. This was done in ~\cite{Arnowitt:1992aq} (see also ~\cite{Ross:1992tz}).
Subsequent to the works of  ~\cite{Arnowitt:1992aq, Ross:1992tz} there were several analyses 
along these lines (see, e.g., \cite{Nath:1992uda, Nath:1992ty}).
Currently such RG analyses are the standard procedure in generating the mass spectra in supergravity unified models. 

{\it Proton decay:}
It was pointed out by Weinberg~\cite{Weinberg:1981wj} and by  Sakai and Yanagida~\cite{Sakai:1981pk}
that 
the supersymmetric grand unified models 
contain baryon and lepton number violating dimension five operators.  Initial investigation of the main decay
modes of the supersymmetric GUTs was done in ~\cite{Dimopoulos:1981dw,Ellis:1981tv}.
 The first analysis within supergravity united model was carried out 
in ~\cite{Arnowitt:1985iy,Nath:1985ub}.
  The supergravity analysis 
 was extended in  several further works: \cite{Nath:1988tx,Arnowitt:1989ud,Nath:1992uda,Nath:1992ty,Arnowitt:1998uz,Nath:2000tr,Arnowitt:1993pd,Nath:1997jc,Arnowitt:1987hx,Arnowitt:1987xk} 
(for the current status of proton decay vs experiment  see ~\cite{Nath:2006ut,Babu:2013jba}).\\
{\it Dark matter:}
Soon after the formulation of supergravity grand unification it was observed 
\cite{Goldberg:1983nd,Krauss:1983ik,Ellis:1983ew}
that with R parity conservation that the neutralino could be a candidate for dark matter.
Later it was shown by computation of the sparticle spectrum using RG evolution that under 
color and charge conservation that the neutralino was indeed the lightest supersymmetric particle
and being neutral it was in fact a viable candidate for dark matter~\cite{Arnowitt:1992aq}.
The precision computation of  relic density using integration over the poles in the annihilation of neutralinos
was given in \cite{Nath:1992ty,Arnowitt:1993mg} using the technique used previously in   the analysis of
non-supersymmetric dark matter analyses~\cite{Griest:1990kh}.
Later on Dick and I worked on the direct detection of dark matter ~\cite{Arnowitt:1994dh,Nath:1994ci,Nath:1994tn,Nath:1994ci,Arnowitt:1994dh,Arnowitt:1995vg,Nath:1997qm,Arnowitt:1998uz}. An analysis of the annual  modulation effect 
on event rates in dark matter detectors  was carried out in 
\cite{Arnowitt:1999gq}.   Dick continued the work on dark matter with other colleagues in later years
(see e.g.,    \cite{0802.2968}).  \\

{\it String inspired supergravity models:}
Since supergravity is the field point limit of strings, the string inspired supergravity models present an
interesting class of high scale models for investigation. A number of models were investigated in
\cite{Nath:1988xn,Arnowitt:1989ur,Nath:1989jr,Nath:1990mb,Arnowitt:1990zw,Nath:1991rk,Arnowitt:1988ax}.
Some of the phenomena  investigated included  $\mu\to e\gamma$ \cite{Arnowitt:1990ww}, 
charged lepton and neutrino masses and mixings~\cite{Arnowitt:1990zw}, and Higgs boson 
phenomenology~\cite{Nath:1991rk}. In~\cite{Wu:1990ay}
Yukawa couplings were computed for the  model $CP^3\times CP^2/Z_3\times Z_3'$.
 In  a later work detecting physics in the post GUT and string scales was carried out ~\cite{Arnowitt:1997ui}.
 
{\it Current status of SUGRA GUTs:}
The discovery of the Higgs boson at $126$ GeV has given strong support for supergravity grand unification.
Thus within the Standard Model vacuum stability is not guaranteed beyond scales of $Q\sim 10^{11}$ GeV
while in supergravity grand unification one can allow for the stability of the vacuum up to GUT scales and beyond.
Further, SUGRA GUT models predict the Higgs boson mass to lie below $\sim 130$ GeV~\cite{Akula:2011aa}
and it is quite remarkable 
that the observed value of the Higgs mass obeys this upper limit giving support to the idea of supergravity grand
unification. The  Higgs mass of $\sim 126$ GeV leads to the average SUSY scale to lie in the TeV region which in
part explains the non-observation  of the sparticles  thus far. Other virtues of the SUGRA GUT model include
an explanation of how the electroweak symmetry breaks, i.e., it breaks via renormalization group effects.
Thus it solves a major problem of the 
standard  model where the Higgs mass is assumed to be tachyonic in an ad hoc fashion.
 It should be noted that historically SUGRA GUT provided the first hint that the top quark should be heavy, i.e., have a mass greater than  $\sim 100$ GeV~\cite{AlvarezGaume:1983gj}.
 Further, in SUGRA GUT one can show 
that with charge and color conservation that the lightest supersymmetric particle is the neutralino over most of the
parameter space of the model and thus the neutralino is a possible candidate for dark matter under the assumption
of R party conservation. Detailed analyses show that the relic density of neutralinos consistent with the WMAP 
~\cite{Hinshaw:2012aka} and Planck ~\cite{Ade:2013ktc}
data
can be gotten. Additionally, the current dark matter experiments are probing the parameter space of supergravity
unified models in the neutralino$-$proton cross sections vs the neutralino mass plane and future detectors
such as XENON1T ~\cite{Aprile:2012zx} and  LUX-ZEPLIN ~\cite{Ghag:2014uva} 
can test a very significant part of the parameter space of SUGRA models. 
The measurement of the Higgs boson mass at 126 GeV points to a  SUSY mass scale in the TeV region.
Such a mass scale provides the desired loop correction needed to lift the tree level Higgs mass to the 
experimentally measured value. It helps suppress the flavor changing neutral current processes and 
also helps stabilize the proton against decay via B\&L violating dimension five operators~\cite{Liu:2013ula}.
The RUN-II of the LHC will significantly expand the region of the parameter space of SUGRA models that will be probed~\cite{Baer:2009dn}. It is hoped that the new data expected from   LHC Run-II will provide us further evidence for the validity of supersymmetry and for SUGRA grand unification (For recent developments in SUGRA GUTs see, \cite{Francescone:2014pza,Nath:2015dza}).

The memory of Richard Arnowitt will live on through his many contributions to physics.
He will also live on in the memory of those who were privileged to know him.
The Memorial Symposium at Texas A\&M, College Station,  September 19-20 was a fitting  tribute to the life and times of Richard
Arnowitt.  Thanks to Marlan Scully and Roland Allen  and the physics department at Texas A\&M 
for organizing the Memorial Symposium.\\~\\

\begin{acknowledgments}
PN's research  is  supported in part by the NSF grant PHY-1314774.
\end{acknowledgments}

\end{document}